\documentclass{elsart}
\usepackage[dvips]{graphicx}

\newcommand{\CP}{\mbox{\em CP}}
\newcommand{\T}{\mbox{\em T}}
\newcounter{formula}
\newcounter{forline}
\newcounter{asym}
\begin{document}

\begin{frontmatter}

\title{
How Could \CP-Invariance and Physics Beyond SM Be Tested in Polarized
Proton Collisions at RHIC?
}

\author{V.~L.~Rykov\thanksref{CA}}
\address{Department of Physics and Astronomy, Wayne State University,
\\ Detroit, MI 48201, USA}
\thanks[CA]{Phone: (313)--577--2781; fax: (313)--577--0711;
e--mail: rykov@physics.wayne.edu}

\begin{abstract}
   Just in months ahead, the first high luminosity collisions of two
polarized proton beams are expected to occur at RHIC in
BNL at $\sqrt{s}$\, up to 500~GeV~\cite{bunce:92}, bringing a new
quality to the collider physics. In collisions of polarized particles,
the presence of two axial vectors of initial polarizations, fully
controlled by experimenters, may dramatically increase the number of
available for tests correlations between participating vectors,
generating asymmetries that could relatively easily be measured. In
frame of Standard Model (SM), many of these asymmetries are either
strongly suppressed or strictly prohibited. Therefore, if some of them
were found nonzero, this could be an indication of a new physics
beyond SM. If certain criteria met, it might be difficult to explain
the observed nonzero correlations in theories without \CP- and/or
\T-violation.
\end{abstract}

\end{frontmatter}

\section*{Introduction}

   The expectations for a New Physics at the energy scale of hundreds
of GeV are high. To some extent, the current status of SM is
reminiscent to that of weak interactions in early 70th just before
the discovery of $J/\psi$-meson and $c$-quark. At that time, something
new had been expected to happen at the scale of a few GeV. Otherwise,
the theories of weak interactions mediated by a vector $W$-boson could
not survive because of their failure to explain the low
$K_{L}^{0}\rightarrow\mu^{+}\mu^{-}$\, decay
rate~\cite{salam:64}. Since 1977~\cite{veltman:77}, it
is well understood that SM of electroweak interactions would
experience quite serious complications if some new physics (Higgs?
Compositeness? \ldots) does not show up at the energies below
$\sim$1~TeV. The energy range just above or even about $W$- and
$Z$-masses is not excluded~\cite{erler:98}.

   For 35 years since the discovery of \CP-violation~\cite{cronin:64},
many attempts have been undertaken to find more \CP- and/or
\T-violating processes other than few known nonleptonic and
semileptonic $K^{0}$-decays. The negative results of these searches
have found a quite natural explanation in frame of
Ca\-bibbo-Ko\-ba\-ya\-shi-Mas\-ka\-wa (CKM) quark mixing matrix
phenomenology with a single nonzero phase that makes this matrix
complex~\cite{ckm:73}. At present time, the CKM-matrix formalism is
{\em de facto\/} the common language for SM interpretation of all
already discovered \CP-violating phenomena. At the currently achieved
level of experimental and theoretical uncertainties, no clear
contradictions had been found of the relations between few measured
\CP-odd parameters and the constrains of CKM-matrix.

   A noticeably high ``\CP-activity'' has been quite evident for the last
decade, and there is even more to come. Plans for \CP-violation studies
are in the research programs of all particle physics accelerator
laboratories around the world, not mentioning {\em B}-factories
proposed almost exclusively with the \CP-violation in mind. Not
undermining in any way the indisputable significance of the recent
\CP-activity, it should be underlined that its efforts are focused mainly
on testing the CKM-based SM predictions for ``direct'' \CP-violation
in $K^{0}$-decays as well as large \CP-violation in $b$-quark
transformations. Meanwhile, there is a quite widespread
dissatisfaction and
disbelief~\cite{dolgov:92,derujula:91,paschos:96,gastoldi:89,garisto:91,kane:92,im:93}
that a single nonzero phase in CKM-matrix may be the only emergence of
\CP-violation in foreseeable energy range, and there are good reasons
for such a discomfort. For example, various approaches to
understanding the baryon asymmetry of the Universe require much
stronger \CP-violation than it is suggested in the minimal CKM-based
SM~\cite{dolgov:92}. In many models, \CP-odd effects, sufficiently
large for being detectable at high energy colliders, could be
generated~\cite{dolgov:92,derujula:91,kane:92,im:93,tev2000:96}. On
the other hand, as it had been pointed out just a few years
ago~\cite{im:93}: {\em ``At present time published limits do not exist
for the size of most CP-violating processes at 100~GeV scale. Thus
heretofore undetected large ($\sim$50\%) CP-violation could occur in
some processes at high energy hadron colliders''.} 

   Since time of the first $p\overline{p}$-collisions in the SPS,
the energy scale of $\sim$10$^{2}$~GeV is a common playground for
experimenters. For the last decade, a number of papers have been
published\footnote{See, for example,
refs.~\cite{kane:92,im:93,tev2000:96,handoko:98}.}, exploring the
feasibility to test \CP- and \T-invariance at large colliders in the
modes other than $B$-decays. Most publications are focused on
unpolarized colliders where the detection of \CP- and/or \T-violation
requires to carry out quite difficult measurements of either fine
balances between particle and antiparticle production in \CP-conjugate
processes, or polarizations\footnote{\ldots or
handedness~\cite{efremov:92,efremov:98} \ldots} of final jets to
detect potentially nonzero \T-odd correlations. Recently, some
indications of the jet handedness correlations in $Z^{0}$-decays,
which were of the opposite sign to that expected in SM, had been
reported~\cite{efremov:98}. The authors looked into the jet
fragmentation mechanisms for a possible explanation. Alternatively, a
presence of a tensor $q\overline{q}Z$-coupling may also be responsible
for the wrong sign correlations.

   In the collisions of polarized particles, the number of potentially
interesting \T-odd as well as \T-even
correlations\footnote{\CP-violation may show up not only via \T-odd
asymmetries but also through purely \T-even ones (see
secs.~\ref{sec:model} and \ref{sec:discussion} for examples).} built
from two axial\footnote{Polarization vectors of projectiles.} and many
polar\footnote{Momenta of incident and final particles and jets in
various processes.} vectors could be enormously large. A new physics
and \CP/\T-violation mechanism behind these correlations could
virtually be anywhere and
everywhere~\cite{dolgov:92,derujula:91,paschos:96,gastoldi:89,garisto:91,kane:92,im:93,tev2000:96,handoko:98,tdlee:73}:
in QCD quark-gluon and gluon-gluon coupling; in electroweak
interactions mediated by the usual vector bosons as well as by
suggested in some theories $W^{\prime}$\, and $Z^{\prime}$\, of higher
masses; in Higgs sector; in the leptoquark exchange; due to
spontaneous \CP/\T-violation; etc. However, an observation of a
nonzero \T-odd correlation may never be treated by itself as an
evidence of \T- or \CP-violation because these correlations are easily
generated by purely \CP- and \T-even initial and final state
interactions. This is particularly serious issue at low momentum
transfers, resulting in a very limited number of low energy processes
being suitable to really test \CP/\T-invariance by measuring \T-odd
correlations~\cite{garisto:91,sakurai:58,burgy:60,sehgal:92}.
At momentum transfers $\sim$10$^{4}$~(GeV/c)$^{2}$ and more, the
``spurious'' asymmetries due to initial and final state interactions
are expected to be small~\cite{kane:92,im:93}. Nevertheless, even in
high energy hard collisions, a comparison of asymmetries in
\CP-conjugate processes is still required for a conclusion on
unambiguous and model independent observation of \CP-violation,
i.e. the same problem as in the measurements of balances between
particle and antiparticle productions needs to be solved.

   With all similarities, there is one important difference between
comparisons of production cross sections and spin dependent
asymmetries in \CP-conjugate processes. In the case of cross sections,
the \CP-noninvariance of a detector may be the cause of false
\CP-odd-like effects due to unaccounted differences in detection
efficiencies to particles and antiparticles\footnote{See discussion in
ref.~\cite{im:93}.}. In contrast, the measurement of spin dependent
asymmetry in any particular process is usually not sensitive to
uncertainties of the overall detection efficiency scale. Therefore,
unaccounted differences of these scales in \CP-conjugate processes do
not introduce systematic errors to the asymmetry comparison. It is
also worth noticing that, in the cases of small background (spurious)
asymmetries, just ``wrong'' relative signs of detected asymmetries in
\CP-conjugate processes\footnote{i.e. inconsistent with the
\CP-invariance; production cross sections could never be of ``wrong''
relative signs.} would be a sufficient evidence for
\CP-violation. And the last not the least, most reasonable models with
\CP-violation quite easily generate \CP-odd spin dependent
asymmetries, while measurements of cross section differences, averaged
over initial and final polarizations, are often not sensitive to
\CP-violating amplitudes. The number of asymmetries to test would be
particularly large if both transversely and longitudinally polarized
projectiles were available.

   Unlike in the accelerator experiments focused on decays of
secondaries, the primary collisions and incident polarized particles
themselves are parts of the reactions under investigation in the
measurements of spin dependent asymmetries. Therefore, for having such
experiments on \CP-violation to be conclusive, an availability of
\CP-conjugate initial states is a must. From this point of view,
$p\overline{p}$-colliders with two polarized
beams~\cite{rossmanith:88} would clearly be among the best and the
most capable machines. In polarized $p\overline{p}$\, collision, all
variety of quark-antiquark, quark-gluon, and gluon-gluon interactions
in \CP-conjugate processes could be tested. Complimentary, polarized
$e^{+}e^{-}$- and $\mu^{+}\mu^{-}$-colliders could be used for
studying \CP/\T-violation in the neutral current lepton scattering and
annihilation, and at the energies above the threshold of $W$-pair
production, the \CP/\T-invariance of charged current interactions
could also be checked. Unfortunately, neither of these colliders with
two polarized beams is currently available. Meanwhile, the
\CP-asymmetric collisions of two polarized proton beams are expected
to be seen soon at RHIC in BNL~\cite{bunce:92}, pursuing the goal to
comprehensively explore the proton spin structure~\cite{makdisi:97}
as well as carry out a wide range study of parity violating phenomena
in $W^{\pm}$- and $Z^{0}$-boson productions and
decays~\cite{bourrely:89}. Although being obviously not the
best for \CP-invariance tests, collisions of two high energy polarized
protons still have capabilities for, at minimum, scouting the problem
in the processes with a reasonably clear picture of underlining parton
interactions. At RHIC, the processes of hadronic leptoproduction via
photon, $W^{\pm}$-, and $Z^{0}$-boson exchange,
\renewcommand{\theequation}{\theformula}
\stepcounter{formula}
\begin{equation}
q\overline{q} \rightarrow W^{\pm} \rightarrow l\overline{l}
\mbox{\hspace*{1.0cm} and \hspace*{1.0cm}}
q\overline{q} \rightarrow Z^{0}/\gamma \rightarrow l^{+}l^{-}\;,
\label{eq:qq_wz_ll}
\end{equation}
will probably be the cleanest ones to search for \CP- and/or
\T-violation and other unusual and unexpected phenomena in.

In the rest of this paper, few examples of measurable asymmetries
for processes~(\ref{eq:qq_wz_ll}) are shown\footnote{Some double-spin
asymmetries had briefly been discussed earlier in
reports~\cite{rykov:98}.}. These asymmetries, if found nonzero, could
be an indication of a new physics beyond SM and, if certain criteria
met, of a \CP- and/or \T-violation.

\section{Phenomenological model (example)}
\label{sec:model}

\begin{figure}[htb]
\centerline{\hbox{
\includegraphics[width=14.0cm,bb=130 590 510 690]{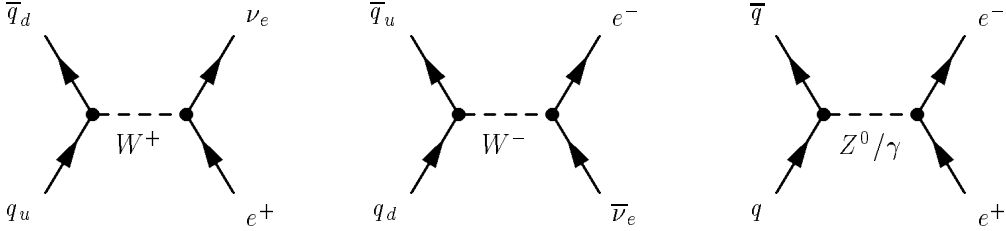}
}}
\caption[]
{
\footnotesize
The lowest order graphs for $q\overline{q} \rightarrow l\overline{l}$ 
($e^{+}/e^{-}$\, and $\nu_{e}/\overline{\nu}_{e}$\, are shown here for
leptons).
}
\label{fig:production}
\end{figure}

   In the lowest order, processes (\ref{eq:qq_wz_ll}) are represented
by the $s$-channel annihilation graphs shown in
Fig.~\ref{fig:production}. Unusual interactions, including \CP- and/or
\T-violation, may be present in either or both of two vertices. To be
specific, the further discussion is held in frame of one
phenomenological modification of SM's electroweak coupling which
had earlier been used elsewhere~\cite{kane:92}. In this model, an
effective Lagrangian of charged  current $q_{u}q_{d}W$-interactions
is:
\stepcounter{formula}
\begin{eqnarray}
L_{c} & = &
\frac{g}{2\sqrt{2}}\Big\{
\big[W_{\mu}^{-}\overline{q}_{d}\gamma^{\mu}
(f_{V}^{+}+f_{V}^{-}\gamma_{5})q_{u} +
W_{\mu}^{+}\overline{q}_{u}\gamma^{\mu}
(f_{V}^{^{*}\!\!+}+f_{V}^{^{*}\!\!-}\gamma_{5})q_{d}\big] +
\nonumber \\
& & +\frac{1}{\Lambda_{q}}
\big[\partial_{\nu}W_{\mu}^{-}\overline{q}_{d}\sigma^{\mu\nu}
(f_{T}^{+}+f_{T}^{-}\gamma_{5})q_{u} +
\partial_{\nu}W_{\mu}^{+}\overline{q}_{u}\sigma^{\mu\nu}
(f_{T}^{^{*}\!\!+}-f_{T}^{^{*}\!\!-}\gamma_{5})q_{d}\big]\Big\}
\label{eq:lagrangian}
\end{eqnarray}
where $g$\, is a coupling constant, presumably on the order of the
electroweak one, $\Lambda_{q}$\, is the energy scale of the ``full
strength'' tensor\footnote{The shortcuts ``vector'' and ``tensor''
are used for couplings without and with derivatives $\partial_{\nu}$,
respectively.} interactions, and the asterisk denotes the complex
conjugate. The notations $q_{u}$\, and $q_{d}$\, are for the ``upper''
($u$\, and $c$)\footnote{$t$-quark is virtually not reachable at RHIC.}
and ``lower'' ($d, s, b$) quarks respectively. The usual
($V$--$A$) interactions correspond to $f_{V}^{+}=f_{V}^{-}=1$\, with
$f_{T}^{\pm}$\, equal to zero\footnote{In this paper: $\gamma_{5}=
-i\gamma^{0}\gamma^{1}\gamma^{2}\gamma^{3}$\, and $\sigma^{\mu\nu} =
\frac{1}{2}(\gamma^{\mu}\gamma^{\nu} - \gamma^{\nu}\gamma^{\mu})$.}.
The \CP- and \T-invariance of model (\ref{eq:lagrangian}) is broken if
any or all ``formfactors'' $f_{V,T}^{\pm}$\, were complex. In the
neutral current effective Lagrangian of type (\ref{eq:lagrangian}) for
$q\overline{q}Z$-interactions\footnote{\ldots of a truly neutral
$q\overline{q}$-pair, consisting of a fermion $q$\, and
\underline{its} antiparticle.}, three formfactors, $f_{V}^{\pm}$\, and
$f_{T}^{+}$, must be real. \CP-violation may occur due to only tensor
coupling with a purely imaginary $f_{T}^{-} \neq 0$. In the
calculations below, the effective Lagrangians of type
(\ref{eq:lagrangian}) with changed notations:
$f_{V,T}^{\pm}\rightarrow l_{V,T}^{\pm}$\, and
$\Lambda_{q}\rightarrow\Lambda_{l}$, have also been used to describe
the lepton coupling to $W$- and $Z$-bosons. In this phenomenological
and rather illustrative example, we do not speculate on the nature of
underlining fundamental interactions which may potentially induce
nonstandard terms and \CP-odd phases in Lagrangian
(\ref{eq:lagrangian}). We do not speculate on the expected magnitudes
for these terms either. We just mention here that the currently
available experimental data from the low energy searches for a weak
tensor coupling generally do not exclude $f_{T}^{\pm}$\, and/or
$l_{T}^{\pm} \sim$1 for
$\Lambda_{q,l}~\geq$~10$^{2}$~GeV~\cite{commins:83}.

   In model (\ref{eq:lagrangian}), the tree-level cross section for
charged current annihilation (\ref{eq:qq_wz_ll}) of polarized quark
and antiquark could be written as:
\stepcounter{formula}
\begin{equation}
\frac{\d\hat{\sigma}}{\d\Omega} \simeq
\frac{g^{4}\hat{s}\big\{\mid M_{fi}^{VV}\mid^{2}+
\frac{\sqrt{\hat{s}}}{\Lambda_{q}}\mid M_{fi}^{VT}\mid^{2}+
\frac{\hat{s}}{\Lambda_{q}^{2}}\mid T_{fi}^{TT}\mid^{2}\big\}}
{64\big\{(\hat{s}-M_{W}^{2})^{2}+
\hat{s}^{2}\Gamma_{W}^{2}/M_{W}^{2}\big\}}
\label{eq:cross_section}
\end{equation}
where $M_{W}$\, and $\Gamma_{W}$\, are for the $W$-boson mass and
width, respectively; $\sqrt{\hat{s}}$\, is the total center-of-mass
(c.m.) energy of colliding quark and antiquark; $\mid
M_{fi}^{VV}\mid^{2}$\, represents the contribution of
$q\overline{q}W$\, vector coupling; $\mid M_{fi}^{TT}\mid^{2}$\, is
due to $q\overline{q}W$\, tensor coupling; and $\mid
M_{fi}^{VT}\mid^{2}$\, is for the interference between vector and
tensor. Using the standard technique, the following formulae for $\mid
M_{fi}\mid^{2}$, summed over final lepton polarizations, could be 
obtained in the c.m. frame of colliding quark and
antiquark\footnote{There are no doubts that various pieces of these
formulae were published and known since, probably, 60th--70th and some
even earlier.}:
\begin{flushleft}
\small
\stepcounter{formula}
\setcounter{asym}{\theformula}
\renewcommand{\theequation}{\theformula.\theforline}
\begin{eqnarray}
\mid M_{fi}^{VV}\mid^{2} & = & \frac{1}{16}
\Big\{\mid f_{V}^{+}+f_{V}^{-}\mid^{2}\cdot\mid
l_{V}^{+}+l_{V}^{-}\mid^{2}(1-\lambda_{q})(1+\lambda_{\overline{q}})
+ \nonumber \\
& & + \mid f_{V}^{+}-f_{V}^{-}\mid^{2}\cdot\mid
l_{V}^{+}-l_{V}^{-}\mid^{2}(1+\lambda_{q})(1-\lambda_{\overline{q}})\Big\}\times
\Big\{1+\mbox{\boldmath $n_{p}\cdot n_{k}$}\Big\}^{2} + \nonumber \\
& + &
\frac{1}{16}\Big\{\mid f_{V}^{+}+f_{V}^{-}\mid^{2}\cdot\mid
l_{V}^{+}-l_{V}^{-}\mid^{2}(1-\lambda_{q})(1+\lambda_{\overline{q}})
+ \nonumber \\
& & + \mid f_{V}^{+}-f_{V}^{-}\mid^{2}\cdot\mid 
l_{V}^{+}+l_{V}^{-}\mid^{2}(1+\lambda_{q})(1-\lambda_{\overline{q}})\Big\}\times
\Big\{1-\mbox{\boldmath $n_{p}\cdot n_{k}$}\Big\}^{2} + \nonumber \\
& + & \frac{\hat{s}}{8\Lambda_{l}}
\Big\{\mid
f_{V}^{+}+f_{V}^{-}\mid^{2}(1-\lambda_{q})(1+\lambda_{\overline{q}}) +
\mid
f_{V}^{+}-f_{V}^{-}\mid^{2}(1+\lambda_{q})(1-\lambda_{\overline{q}})\Big\}\times
\nonumber \\
& & \times\Big\{\mid l_{T}^{+}\mid^{2}+\mid l_{T}^{-}\mid^{2}\Big\}\times
\Big\{1-\mbox{\boldmath $(n_{p}\cdot n_{k})$}^{2}\Big\} +
\stepcounter{forline}
\label{eq:vv_long} \\
& + & \frac{1}{2}\Big\{(\mid l_{V}^{+}\mid^{2}+\mid l_{V}^{-}\mid^{2}) -
\frac{\hat{s}}{\Lambda_{l}^{2}}(\mid l_{T}^{+}\mid^{2}+\mid
l_{T}^{-}\mid^{2})\Big\}\times
\Big\{(\mid f_{V}^{+}\mid^{2}-\mid f_{V}^{-}\mid^{2})\times
\nonumber \\
& & \times\{\mbox{\boldmath $(\zeta^{\perp}_{q}\cdot n_{k})$}
\mbox{\boldmath $(\zeta^{\perp}_{\overline{q}}\cdot n_{k})$} -
\frac{1}{2}\mbox{\boldmath
$(\zeta^{\perp}_{q}\cdot\zeta^{\perp}_{\overline{q}})$} 
[1-\mbox{\boldmath $(n_{p}\cdot n_{k})$}^{2}]\} \pm
\stepcounter{forline}
\label{eq:vv_teven} \\
& & \pm\mathop{\mathrm{Im}}(f^{+}_{V}f^{^{*}\!\!-}_{V})\times
\nonumber \\
& & \times\{\mbox{\boldmath $(n_{k}\cdot [\zeta^{\perp}_{q}\times
n_{p}])(\zeta^{\perp}_{\overline{q}}\cdot n_{k})$}+
\mbox{\boldmath $(n_{k}\cdot [\zeta^{\perp}_{\overline{q}}\times
n_{p}])(\zeta^{\perp}_{q}\cdot
n_{k})$}\}\Big\}
\stepcounter{forline}
\label{eq:vv_todd}
\end{eqnarray}
\stepcounter{formula}
\setcounter{forline}{0}
\begin{eqnarray}
\mid M_{fi}^{TT}\mid^{2} & = & \frac{1}{8}
\Big\{(\mid l_{V}^{+}\mid^{2} + \mid l_{V}^{-}\mid^{2})
[1-\mbox{\boldmath $(n_{p}\cdot n_{k})$}^{2}] +
\frac{\hat{s}}{\Lambda_{l}^{2}}
(\mid l_{T}^{+}\mid^{2} + \mid l_{T}^{-}\mid^{2})\mbox{\boldmath
$(n_{p}\cdot n_{k})$}^{2}\Big\}\times
\nonumber \\
& \times & \Big\{\mid f_{T}^{+}\pm
f_{T}^{-}\mid^{2}(1-\lambda_{q})(1-\lambda_{\overline{q}})+
\mid f_{T}^{+}\mp
f_{T}^{-}\mid^{2}(1+\lambda_{q})(1+\lambda_{\overline{q}})+
\stepcounter{forline}
\label{eq:tt_long} \\
& & + 2 (\mid f_{T}^{+}\mid^{2}-\mid f_{T}^{-}\mid^{2})
\mbox{\boldmath $(\zeta^{\perp}_{q}\cdot\zeta^{\perp}_{\overline{q}})$}-
4 \mathop{\mathrm{Im}}(f^{+}_{T}f^{^{*}\!\!-}_{T})
\mbox{\boldmath
$(n_{p}\cdot[\zeta^{\perp}_{q}\times\zeta^{\perp}_{\overline{q}}])$}\Big\}
\stepcounter{forline}
\label{eq:tt_trans}
\end{eqnarray}
\renewcommand{\theequation}{\theformula}
\stepcounter{formula}
\begin{eqnarray}
\mid M_{fi}^{VT}\mid^{2} & = & \frac{1}{2}\Big\{
(\mid l_{V}^{+}\mid^{2}+\mid l_{V}^{-}\mid^{2}) -
\frac{\hat{s}}{\Lambda_{l}^{2}}
(\mid l_{T}^{+}\mid^{2}+\mid l_{T}^{-}\mid^{2})\Big\}\times
\mbox{\boldmath $(n_{k}\cdot n_{p})$}\times 
\nonumber \\
& \times & \Big\{\mathop{\mathrm{Re}}[(f^{+}_{V}f^{^{*}\!\!+}_{T} \mp
f^{-}_{V}f^{^{*}\!\!-}_{T}) +
\lambda_{\overline{q}} (\mp f^{+}_{V}f^{^{*}\!\!-}_{T}+
f^{-}_{V}f^{^{*}\!\!+}_{T})]
\mbox{\boldmath $(n_{k}\cdot [\zeta^{\perp}_{q}\times n_{p}])$} -
\nonumber \\ 
& - & \mathop{\mathrm{Im}}[(f^{+}_{V}f^{^{*}\!\!-}_{T} \mp
f^{-}_{V}f^{^{*}\!\!+}_{T}) +
\lambda_{\overline{q}} (\mp f^{+}_{V}f^{^{*}\!\!+}_{T}+
f^{-}_{V}f^{^{*}\!\!-}_{T})] 
\mbox{\boldmath $(\zeta^{\perp}_{q}\cdot n_{k})$} + \nonumber \\
& + & \mathop{\mathrm{Re}}[(f^{+}_{V}f^{^{*}\!\!+}_{T} \pm
f^{-}_{V}f^{^{*}\!\!-}_{T}) -
\lambda_{q} (\pm f^{+}_{V}f^{^{*}\!\!-}_{T} + f^{-}_{V}f^{^{*}\!\!+}_{T})]
\mbox{\boldmath $(n_{k}\cdot [\zeta^{\perp}_{\overline{q}}\times n_{p}])$} +
\nonumber \\
& + & \mathop{\mathrm{Im}}[(f^{+}_{V}f^{^{*}\!\!-}_{T} \pm
f^{-}_{V}f^{^{*}\!\!+}_{T}) -
\lambda_{q} (\pm f^{+}_{V}f^{^{*}\!\!+}_{T} + f^{-}_{V}f^{^{*}\!\!-}_{T})]
\mbox{\boldmath $(\zeta^{\perp}_{\overline{q}}\cdot n_{k})$}\Big\} +
\nonumber \\ 
& + & \mathop{\mathrm{Re}}(l^{+}_{V}l^{^{*}\!\!-}_{V})\times
\nonumber \\
& \times & \Big\{\mathop{\mathrm{Re}}[(\mp f^{+}_{V}f^{^{*}\!\!-}_{T} +
f^{-}_{V}f^{^{*}\!\!+}_{T}) +
\lambda_{\overline{q}} (f^{+}_{V}f^{^{*}\!\!+}_{T}\mp
f^{-}_{V}f^{^{*}\!\!-}_{T})] 
\mbox{\boldmath $(n_{k}\cdot [\zeta^{\perp}_{q}\times n_{p}])$} -
\nonumber \\ 
& - & \mathop{\mathrm{Im}}[(\mp f^{+}_{V}f^{^{*}\!\!+}_{T} +
f^{-}_{V}f^{^{*}\!\!-}_{T}) +
\lambda_{\overline{q}} (f^{+}_{V}f^{^{*}\!\!-}_{T}\mp
f^{-}_{V}f^{^{*}\!\!+}_{T})] 
\mbox{\boldmath $(\zeta^{\perp}_{q}\cdot n_{k})$} + \nonumber \\
& + & \mathop{\mathrm{Re}}[(\pm f^{+}_{V}f^{^{*}\!\!-}_{T} +
f^{-}_{V}f^{^{*}\!\!+}_{T}) -
\lambda_{q} (f^{+}_{V}f^{^{*}\!\!+}_{T}\pm f^{-}_{V}f^{^{*}\!\!-}_{T})]
\mbox{\boldmath $(n_{k}\cdot [\zeta^{\perp}_{\overline{q}}\times
n_{p}])$} + 
\nonumber \\
& + & \mathop{\mathrm{Im}}[(\pm f^{+}_{V}f^{^{*}\!\!+}_{T} +
f^{-}_{V}f^{^{*}\!\!-}_{T}) -
\lambda_{q} (f^{+}_{V}f^{^{*}\!\!-}_{T}\pm f^{-}_{V}f^{^{*}\!\!+}_{T})]
\mbox{\boldmath $(\zeta^{\perp}_{\overline{q}}\cdot n_{k})$}\Big\}
\label{eq:vt_trans}
\end{eqnarray}
\normalsize
\end{flushleft}
In the formulae above, \mbox{\boldmath $n_{p}=p/\!\mid p\mid$} and
\mbox{\boldmath $n_{k}=k/\!\mid k\mid$} where \mbox{\boldmath $p
\equiv p_{q}$} is the momentum of the incident quark and
\mbox{\boldmath $k \equiv k_{e^{-}}$} or \mbox{\boldmath $k\equiv
k_{\nu}$} is the momentum of the final lepton; $\lambda_{q}$ =
\mbox{\boldmath $(\zeta_{q}\cdot p_{q})/\mid p_{q}\mid$} and
$\lambda_{\overline{q}}$ = \mbox{\boldmath $(\zeta_{\overline{q}}\cdot
p_{\overline{q}})$/$\mid p_{\overline{q}}\mid$} are for helicities,
and \mbox{\boldmath $\zeta_{q}^{\perp}$} and \mbox{\boldmath
$\zeta_{\overline{q}}^{\perp}$} for transverse polarizations of the
incident quark and antiquark, respectively. Only the main contributions
to the cross section were kept in
eqs.~(\theasym)--(\ref{eq:vt_trans}), neglecting all others
suppressed by powers of $m_{q,l}/\sqrt{\hat{s}}$, where $m_{q,l}$\,
is for the masses of participated in the reaction quarks and leptons.
The upper and lower signs correspond to $W^{+}$\, and $W^{-}$\,
productions, respectively\footnote{For a neutral current process,
the result does not depend on these signs due to restrictions on
allowed choices for $f_{V,T}^{\pm}$.}.

\section{Discussion}
\label{sec:discussion}

   As one can see from
eqs.~(\ref{eq:cross_section})--(\ref{eq:vt_trans}), new physics in the
processes (\ref{eq:qq_wz_ll}) at polarized hadron colliders may show
up via a number of measurable characteristics.

   The deviation of the $q\overline{q}W$-coupling from the standard
($V$--$A$) form\footnote{\ldots or from the usual SM's mixture of {\em
V\/} and {\em A\/} in the neutral current
$q\overline{q}Z/\gamma$-coupling.} affects the lepton production
distributions in polar angle $\theta =$ \mbox{\boldmath
$\widehat{k\,p}$}. However, this could hardly be exploited for new
physics searches at hadron colliders because $\theta$-distributions
are also products of not so well known parton structure functions. In
the real life, the measurements of lepton spectra and
$\theta$-distributions are rather used to reconstruct structure
functions, assuming the known interactions at the parton level.

   The same is true for dependences~(\ref{eq:vv_long})
and~(\ref{eq:tt_long}) of the production cross sections on quark
helicities. To reconstruct these dependences back to the quark-parton
level, using experimental data from hadron collisions, the knowledge of
the spin-dependent parton structure functions is required. In the real
life, again, the situation is quite opposite: the measurements of these
structure functions themselves will be the main focus of the RHIC
experimental program with polarized protons. At the tree-level of the
model in consideration, neither $\theta$-distributions nor helicity
dependences of cross sections are sensitive to the \CP-violating
phases in Lagrangian~(\ref{eq:lagrangian}), although this does not
mean that such a sensitivity may not be present in others, more
sophisticated theories.

   The situation with asymmetries (\ref{eq:vv_teven}),
(\ref{eq:vv_todd}), (\ref{eq:tt_trans}), and (\ref{eq:vt_trans}) in
collisions of transversely polarized nucleons is quite different. In
SM, most transverse asymmetries are either strongly suppressed or
strictly prohibited. Therefore, if some of them were found to be
sufficiently large, this would be an indication of deviations from SM,
regardless of the spin structure of colliding hadrons.

   The first example is the well known
{\em P}- and \T-even double-spin azimuthal
anisotropy~(\ref{eq:vv_teven}) of lepton production:
\stepcounter{formula}
\begin{equation}
A_{TT} \propto (\mid f_{V}^{+}\mid^{2}-\mid
f_{V}^{-}\mid^{2})\cdot\mbox{\boldmath
$\mid\zeta_{q}^{\perp}\mid\cdot\mid\zeta_{\overline{q}}^{\perp}\mid$}\cdot\cos
(\varphi_{\overline{q}q}-2\varphi_{kq})\;,
\label{eq:a_tt}
\end{equation}
where
$\varphi_{\overline{q}q} = \widehat{\mbox{\boldmath
$\zeta_{\overline{q}}^{\perp}\,\zeta_{q}^{\perp}$}}$\, and
$\varphi_{kq} = \widehat{\mbox{\boldmath
$k_{\perp}\,\zeta_{q}^{\perp}$}}$\, with \mbox{\boldmath $k_{\perp} =
k-(k\cdot n_{p})n_{p}$}. The tree-level $A_{TT}$~=~0 in charged
current ($V$--$A$)-interactions. However, $A_{TT}$\, is expected to be
measurably large in the neutral current Drell-Yan process,
$q\overline{q}\rightarrow Z^{0}/\gamma\rightarrow l^{+}l^{-}$, making
it attractive for measurements of the quark transversity distributions
$h_{1}(x)$\, in proton~\cite{ralston:79,soffer:95,martin:98}.

   The double-spin \T-odd asymmetry~(\ref{eq:vv_todd}) due to
\CP-violation in the $q\overline{q}W$\, vector coupling looks similar
to $A_{TT}$, but rotated by 45$^{o}$:
\stepcounter{formula}
\begin{equation}
A_{TT}^{\perp}\propto\pm\mathop{\mathrm{Im}}(f^{+}_{V}f^{^{*}\!\!-}_{V})\cdot\mbox{\boldmath
$\mid\zeta_{q}^{\perp}\mid\cdot\mid\zeta_{\overline{q}}^{\perp}\mid$}\cdot\sin
(\varphi_{\overline{q}q}-2\varphi_{kq})\;.
\label{eq:a_tt_perp}
\end{equation}
In model (\ref{eq:lagrangian}), this asymmetry is the ``true'' \CP-odd
one and, because of that, it is of the opposite signs in two
\CP-conjugate processes of $W^{\pm}$-production. A ``spurious''
\CP-even $A_{TT}^{\perp}$\, would not change sign under
\CP-conjugation.
\addtocounter{footnote}{-4}
If $A_{TT}^{\perp}$\, anisotropy would have ever been detected in a
neutral current annihilation\footnotemark, it could never be a true
\CP-odd but only a spurious \CP-even one due to some \CP-conserving
initial and/or final state interactions.
\addtocounter{footnote}{3}

   Without an interference to the vector, the tensor coupling of
transversely polarized quarks to gauge bosons does not generate
azimuthal anisotropies of lepton production at the tree level.
However, it may generate cross section dependences (\ref{eq:tt_trans})
on the relative orientation of vectors \mbox{\boldmath
$\zeta_{q}^{\perp}$} and \mbox{\boldmath
$\zeta_{\overline{q}}^{\perp}$}. The \CP-even cross section difference
$\Delta\sigma_{N}\propto (\mid f_{T}^{+}\mid^{2}-\mid
f_{T}^{+}\mid^{2})\mbox{\boldmath
$(\zeta_{q}^{\perp}\cdot\zeta_{\overline{q}}^{\perp})$}$\, would be
nonzero if $\mid f_{T}^{+}\mid\neq\mid f_{T}^{-}\mid$. \CP-violation
in the tensor coupling makes cross sections dependent on also the
\T-odd product \mbox{\boldmath $(n_{p}\cdot
[\zeta_{q}^{\perp}\times\zeta_{\overline{q}}^{\perp}])$} with the
measurable
$\Delta\sigma_{\perp}=\sigma_{\perp}^{R}-\sigma_{\perp}^{L}\propto\mathop{\mathrm{Im}}(f^{+}_{T}f^{^{*}\!\!-}_{T})\mbox{\boldmath
$(n_{p}\cdot
[\zeta_{q}^{\perp}\times\zeta_{\overline{q}}^{\perp}])$}$, where cross
sections $\sigma_{\perp}^{R,L}$\, are for the ``right''- and
``left''-handed orientations of vectors \mbox{\boldmath
$\zeta_{q}^{\perp}$, $\zeta_{\overline{q}}^{\perp}$, and $n_{p}$},
respectively. In formula~(\ref{eq:tt_trans}), the true \CP-odd
$\Delta\sigma_{\perp}$\, is of the same sign in $W^{+}$- and
$W^{-}$-productions, but a spurious \CP-even $\Delta\sigma_{\perp}$\,
must change its sign under \CP-conjugation. In a neutral current
annihilation, neither \CP-conserving interactions could generate a
nonzero $\Delta\sigma_{\perp}$. Therefore, an observation of
$\Delta\sigma_{\perp}\neq$~0 in a neutral-current quark (or lepton)
and \underline{its} antiparticle annihilation would be an unambiguous
evidence of \CP-violation. With all attractiveness, the reliable
measurements of $\Delta\sigma_{N,\perp}$\, should be expected to be
significantly more difficult compared to azimuthal anisotropies. The
reason is the high sensitivity of
$\Delta\sigma_{N,\perp}$\, measurements to spin misalignments, namely,
to unaccounted longitudinal components of the colliding beam
polarizations.

   The single-spin and double-spin\footnote{\ldots with one quark
transversely and the other one longitudinally polarized \ldots}
asymmetries of type (\ref{eq:vt_trans}), arising from the
vector-tensor interference, are strongly suppressed at the SM's
tree level. Both, \T-even and \T-odd asymmetries may potentially be
generated. As usually, to determine, whether \CP-violation takes
place, the relative signs of detected asymmetries in \CP-conjugate
processes must be compared. It is interesting to notice that, in
eq.~(\ref{eq:vt_trans}), the \CP-odd imaginary parts are before
\T-even correlations \mbox{\boldmath $(\zeta^{\perp}\cdot
n_{k})$}, while the \CP-even real parts of participating formfactors
$f_{V,T}^{\pm}$\, stand before \T-odd products of three vectors, 
\mbox{\boldmath $(n_{k}\cdot [\zeta^{\perp}\times n_{p}])$}. This, to
some extent surprising result, is apparently the feature of the
particular model in consideration, although it is probably not so
difficult to invent other models where \CP-violation could generate
\T-odd single- and double-spin asymmetries of type~(\ref{eq:vt_trans})
as well.

   Imaginary parts of formfactors $l_{V,T}^{\pm}$\, are not present in
formulae~(\theasym)--(\ref{eq:vt_trans}) at all. This means
that, without tracking the polarizations of final leptons,
processes~(\ref{eq:qq_wz_ll}) are not sensitive to \CP/\T-violation in
the lepton sector.

   At the early stage of just ``hunting'' for unusual correlations at
polarized hadron colliders, it is probably not necessary to know much
about the spin-dependent parton structure functions of colliding
hadrons. However, to distinguish \CP-odd and \CP-even correlations by
comparing asymmetries in \CP-conjugate processes, at least the
relative signs of polarizations and often the directions of motions of
the primary hardly collided quark and antiquark need to be known. This
means that it would be necessary, at minimum, to identify with a
reasonably high certainty the parent projectiles of incident quark and
antiquark in every single event of $W^{\pm}$- and/or
$Z^{0}$-production. Then, some knowledges about $h_{1}^{q}$\, and
$h_{1}^{\overline{q}}$\, obtained from, for example, $A_{TT}$\,
measurements in Drell-Yan process could be used to determine the
relative $q$\, and $\overline{q}$\, polarization signs.

   The identification of the predominant parent projectiles for $q$\,
and $\overline{q}$\, would be quite obvious at
$p\overline{p}$-colliders. At proton-proton colliders, where the
situation is  not the same clear, the properties of gauge boson
production kinematics may help. In $pp$-collisions, gauge bosons will
mostly be produced in hard processes, involving a valence quark from
one proton and sea antiquark from the other proton. From multiple
simulations it follows that gauge bosons will likely be moving in the
direction of the incident ``hard'' valence quark rather than in the
direction of a ``soft'' sea antiquark~\cite{saito:95}. This means
that, if the polarization of one beam was altered, the effects
detected in the forward and backward hemispheres in respect to the
direction of this particular beam would have the different origins. In
the forward direction, effects would mainly be due to changed
polarizations of valence quarks. But in the backward hemisphere,
effects would likely be caused by the alteration of polarizations of
sea antiquarks. The production kinematics of $Z$-bosons will fully be
reconstructed in every single event of type~(\ref{eq:qq_wz_ll}) as a
part of the $Z^{0}$-decay identification procedure. In the case of
$W^{\pm}$, the situation is more difficult because of the neutrino
escape. However, according to simulations~\cite{saito:95}, the $W$'s
kinematics could also be determined with some certainty which would
probably be sufficient to separate $W$-bosons, emerging into the
forward and backward hemispheres.

   What might be the expectations for magnitudes of the nonstandard
asymmetries discussed above in polarized proton collisions at RHIC? To
answer this question, two other questions should be addressed
first. In general, question number one is about the predictions of
various theoretical models for the types and sizes of spin dependent
correlations at the quark-parton level. Speaking in narrower terms of
Lagrangian~(\ref{eq:lagrangian}), it is about predictions for the
formfactors $f_{V,T}^{\pm}$\, and $l_{V,T}^{\pm}$\, and their
\CP-violating phases, as well as for the tensor interactions' energy
scales $\Lambda_{q,l}$. We will not discuss these issues here,
leaving them for the expertise of theorists, along with the analysis
of restrictions to parameters of theoretical models, arising from the
available experimental data. The estimations for background
asymmetries due to higher orders SM interactions could also be a
natural part of this expertise.

   Question number two, on how to transfer spin correlations at the
quark-parton level to the asymmetries in proton collisions and back, is
expected to be addressed by simulations in the nearest future. Here, 
just some crude estimates for \CP-odd asymmetries of
model~(\ref{eq:lagrangian}), derived from the $A_{TT}$-study of
ref.~\cite{martin:98}, are provided. In that study, the upper limits
for the $A_{TT}$-asymmetry at RHIC have been obtained, using the
assumption on saturation of Soffer's inequality~\cite{soffer:95} for
the parton transversity distributions in proton. It had been shown that,
for high mass Drell-Yan pairs, the $A_{TT}$-asymmetry in
$pp$-collisions at RHIC could be as large as 3--5\%. The comparison of
the $A_{TT}$-term to terms for the other double-spin asymmetries in
eqs.~(\theasym)--(\ref{eq:vt_trans}) gives that all \CP-odd
double-spin asymmetries should be expected in the same scale of
$\sim$(0.03--0.05) but multiplied, of course, by the factors
$\mathop{\mathrm{Im}}(F_{1}F_{2}^{*})$\, where $F_{1,2}$\, are for
either $f_{V}^{\pm}$\, or $f_{T}^{\pm}\times M_{W}/\Lambda_{q}$,
depending on the particular asymmetry in consideration. The similar
exercise leads to the estimate for single-spin asymmetries at
$\sim\sqrt{\mbox{0.03--0.05}}\times\mathop{\mathrm{Im}}(f_{V}f^{*}_{T})\times
M_{W}/\Lambda_{q}\sim$~(0.1--0.3)$\times\mathop{\mathrm{Im}}(f_{V}f^{*}_{T})\times
M_{W}/\Lambda_{q}$.

   With the statistics accumulation $\sim$(3--4)$\cdot$10$^{3}$\, of
$Z$-boson's and \mbox{$\sim$10$^{4}$--10$^{5}$}\, of $W$'s lepton
decays a year~\cite{derevschikov:95}, RHIC's sensitivity to the spin
asymmetries in processes~(\ref{eq:qq_wz_ll}) is expected to be
$\sim$10$^{-2}$\, at $pp$-level. This makes measurements of the
discussed here double-spin correlations just marginally sensitive to
even large deviations from SM in quark-parton interactions. However, a
sizeable presence of, for example, \CP-even and/or \CP-odd tensor
interactions should be detectable at RHIC through the measurements of
single-spin transverse asymmetries.

\section{Conclusion}
\label{sec:conclusion}

   In this paper, some measurable spin correlations, which may
signal about New Physics beyond SM at RHIC with polarized protons,
have been highlighted. A comparison of these correlation in what is
believed \CP-conjugate processes at the quark-parton interaction level
may provide indications of large \CP- and/or \T-violation at the
energy scale of $\sim$10$^{2}$~GeV. To make it clear, an absolutely
unambiguous and model independent evidence of a \CP-violation could
probably never be obtained from the measurements of spin-dependent
production asymmetries at $pp$-colliders. However, strong
enough indications may potentially emerge if, for example, several
asymmetries have been detected in the same process and, with either
assumption about the spin structure of proton, one part of these
asymmetries behaved as a \CP-even under \CP-conjugation, and the other
part as a \CP-odd, and vice versa. In the situations like this, it
might probably not be so easy to reconcile \CP-conservation with the
current quark-parton picture of gauge boson hadroproduction.

\begin{ack}
   The author is thankful to D.~Boer, G.~Bunce, T.~Cormier,
M.~Grosse-Per\-de\-kamp, R.~Jaffe, Yo.~Makdisi, N.~Saito, N.~Samois,
J.~Soffer, M.~Tannenbaum, and R.~Zul'karneev for the multiple useful
and encouraging discussions. This work was supported in part by the
U.S. Department of Energy Grant DE-FG0292ER40713.
\end{ack}



\end{document}